\documentstyle[12pt]{article}

\catcode`\@=11
\long\def\@makefntext#1{
\protect\noindent \hbox to 3.2pt {\hskip-.9pt
$^{{\ninerm\@thefnmark}}$\hfil}#1\hfill}		

\def\@makefnmark{\hbox to 0pt{$^{\@thefnmark}$\hss}}  

\def\ps@myheadings{\let\@mkboth\@gobbletwo
\def\@oddhead{\hbox{}
\rightmark\hfil\ninerm\thepage}
\def\@oddfoot{}\def\@evenhead{\ninerm\thepage\hfil
\leftmark\hbox{}}\def\@evenfoot{}
\def\sectionmark##1{}\def\subsectionmark##1{}}

\newcounter{sectionc}\newcounter{subsectionc}\newcounter{subsubsectionc}
\renewcommand{\section}[1] {\vspace*{0.6cm}\addtocounter{sectionc}{1}
\setcounter{subsectionc}{0}\setcounter{subsubsectionc}{0}\noindent
	{\normalsize\bf\thesectionc. #1}\par\vspace*{0.4cm}}
\renewcommand{\subsection}[1] {\vspace*{0.6cm}\addtocounter{subsectionc}{1}
	\setcounter{subsubsectionc}{0}\noindent
	{\normalsize\it\thesectionc.\thesubsectionc. #1}\par\vspace*{0.4cm}}
\renewcommand{\subsubsection}[1]
{\vspace*{0.6cm}\addtocounter{subsubsectionc}{1}
	\noindent {\normalsize\rm\thesectionc.\thesubsectionc.\thesubsubsectionc.
	#1}\par\vspace*{0.4cm}}

\newcounter{appendixc}
\newcounter{subappendixc}[appendixc]
\newcounter{subsubappendixc}[subappendixc]

\renewcommand{\appendix}[1] {\vspace*{0.6cm}
        \refstepcounter{appendixc}
        \setcounter{figure}{0}
        \setcounter{table}{0}
        \setcounter{equation}{0}
        \renewcommand{\thefigure}{\Alph{appendixc}.\arabic{figure}}
        \renewcommand{\thetable}{\Alph{appendixc}.\arabic{table}}
        \renewcommand{\theappendixc}{\Alph{appendixc}}
        \renewcommand{\theequation}{\Alph{appendixc}.\arabic{equation}}
        \noindent{\bf Appendix \theappendixc #1}\par\vspace*{0.4cm}}

\def\abstracts#1{{

\centering{\begin{minipage}{12.2truecm}\footnotesize\baselineskip=12pt\noindent
	\centerline{\footnotesize ABSTRACT}\vspace*{0.3cm}
	\parindent=0pt #1
	\end{minipage}}\par}}

\renewenvironment{thebibliography}[1]
	{\begin{list}{\arabic{enumi}.}
	{\usecounter{enumi}\setlength{\parsep}{0pt}
\setlength{\leftmargin 1.25cm}{\rightmargin 0pt}
	 \setlength{\itemsep}{0pt} \settowidth
	{\labelwidth}{#1.}\sloppy}}{\end{list}}

\newcounter{itemlistc}
\newcounter{romanlistc}
\newcounter{alphlistc}
\newcounter{arabiclistc}

\newcommand{\fcaption}[1]{
        \refstepcounter{figure}
        \setbox\@tempboxa = \hbox{\footnotesize Fig.~\thefigure. #1}
        \ifdim \wd\@tempboxa > 6in
           {\begin{center}
        \parbox{6in}{\footnotesize\baselineskip=12pt Fig.~\thefigure. #1}
            \end{center}}
        \else
             {\begin{center}
             {\footnotesize Fig.~\thefigure. #1}
              \end{center}}
        \fi}

\newcommand{\tcaption}[1]{
        \refstepcounter{table}
        \setbox\@tempboxa = \hbox{\footnotesize Table~\thetable. #1}
        \ifdim \wd\@tempboxa > 6in
           {\begin{center}
        \parbox{6in}{\footnotesize\baselineskip=12pt Table~\thetable. #1}
            \end{center}}
        \else
             {\begin{center}
             {\footnotesize Table~\thetable. #1}
              \end{center}}
        \fi}

\def\@citex[#1]#2{\if@filesw\immediate\write\@auxout
	{\string\citation{#2}}\fi
\def\@citea{}\@cite{\@for\@citeb:=#2\do
	{\@citea\def\@citea{,}\@ifundefined
	{b@\@citeb}{{\bf ?}\@warning
	{Citation `\@citeb' on page \thepage \space undefined}}
	{\csname b@\@citeb\endcsname}}}{#1}}

\newif\if@cghi
\def\cite{\@cghitrue\@ifnextchar [{\@tempswatrue
	\@citex}{\@tempswafalse\@citex[]}}
\def\citelow{\@cghifalse\@ifnextchar [{\@tempswatrue
	\@citex}{\@tempswafalse\@citex[]}}
\def\@cite#1#2{{$\null^{#1}$\if@tempswa\typeout
	{IJCGA warning: optional citation argument
	ignored: `#2'} \fi}}

 1
 1
 1

\font\ninerm=cmr9

\newcommand{\beqn}{\begin{equation}}
\newcommand{\eeqn}{\end{equation}}
\newcommand{\barr}{\begin{eqnarray}}
\newcommand{\earr}{\end{eqnarray}}
\begin{document}
\baselineskip=22pt
\centerline{\normalsize\bf DUAL GINZBURG-LANDAU THEORY }
\baselineskip=16pt
\centerline{\normalsize\bf AND CHIRAL SYMMETRY BREAKING}
\vspace{0.8cm}
\begin{center}
{\footnotesize Hiroshi TOKI\footnote{e-mail toki@miho.rcnp.osaka-u.ac.jp},
 Shoichi SASAKI and Hideo SUGANUMA}
\baselineskip=13pt
\centerline{\footnotesize\it RCNP, Osaka University,Ibaraki,Osaka 567,Japan}
\end{center}
\vspace{0.9cm}

\abstracts{We study the properties of quarks, being confined in hadrons,
with the Schwinger-Dyson equation in the dual Ginzburg-Landau Theory.
Magnetic monopole condensation, which provides quark confinement, is
demonstrated responsible also for dynamical chiral-symmetry breaking.
We discuss then the recovery of the chiral symmetry at finite
temperature.}

\normalsize\baselineskip=15pt
\section{Introduction}
Quarks are not found in the free space, while they are seen by deep
inelastic scattering in hadrons. For the hadron physics, it is the
essential problem to understand why and how quarks are confined in hadrons.
Since quark confinement is not understood, there are many popular and
sometimes conflicting models for nucleon suggested in the literature.
In the bag picture, massless quarks are confined within the bag in the
Wigner mode, while quarks are repelled and mesons
are present outside the bag in the Nambu-Goldstone mode.
Depending on the strength of confinement, from weak to strong, we have the
MIT bag model, the chiral bag model and the Skyrmion model, respectively.
In the constituent quark picture, quarks are massive and considered
quasi-particle and they are confined by the long range confining potential.
In this picture, confinement is not sharp but is gradually increasing
with distance. We should understand the mechanism of quark confinement to
settle our picture on nucleons and eventually on entire hadrons.

The chiral symmetry and its spontaneous breaking is also the essential
feature of the hadron physics. In the light quark system, the QCD lagrangian
possesses the chiral symmetry, since we can neglect safely
the small current quark mass as compared to the nucleon mass scale.
If this symmetry is kept, we expect parity doublets in hadron spectra.
This is not the case by observing the meson spectrum as the pion ($0^-$)
has an extremely small mass and the parity partner, the sigma ($0^+$) meson,
is not found, and the rho ($1^-$) and the $A_1$ ($1^+$) mesons do have
completely different masses by about 450MeV.
Confinement of quarks and chiral symmetry breaking are most important
phenomena in the hadron physics.

The dynamics of the hadron physics are believed to be described by QCD.
In the perturbative QCD, the renormalized gauge coupling constant shows the
asymptotic free property at large momentum, $Q$, while it becomes
progressively large at small $Q$, where confinement and chiral symmetry
breaking are expected. What is essential here at small $Q$?

Here, the lattice QCD theory demonstrates quark confinement, chiral
symmetry breaking, hadron masses and many other nonperturbative properties.
The recent simulation studies have shown that the essential degree
of freedom for these phenomena is QCD-monopole and its condensation.
There are many talks on the importance of the QCD monopole field in
Confinement'95.

\section{Dual Ginzburg-Landau Theory}
What is then the QCD-monopoles? 't Hooft has demonstrated that the
non-abelian gauge theory as QCD is reduced to an abelian gauge theory
as QED with magnetic monopoles by abelian gauge fixing, which is a choice
of a particular gauge\cite{hooft}.
Suzuki and his collaborators in the Kanazawa group,
have then discussed extensively the dual Ginzburg-Landau (DGL)
theory\cite{suzuki}.
In the DGL Theory, the non-abelian part is considered negligible and the
QCD-monopole field couples with the dual gauge field. The biggest assumption
is then the introduction of the self interaction term of QCD-monopoles,
which is introduced by hand, to
cause QCD-monopole condensation and then the dual Higgs mechanism,
providing a finite mass, $m_B$, to the dual gauge field.
The details of the DGL theory
are provided in the papers of Matsubara and Suganuma {\it et.al.}
in Confinement'95\cite{{ALL},{matsubara}}.

We can then imitate the dual Meissner effect for quark confinement
as Nambu\cite{nambu} has demonstrated it just after Nielsen and Olesen\cite{NO}
have
formulated the relativistic version of the
Ginzburg-Landau theory on the vortex solution in superconductors.
We only reverse the roles of the electric and the magnetic fields,
which is called {\it dual transformation}, hence the word {\it dual
Ginzburg-Landau
theory}. When the dual gauge field gets a mass, the color electric flux cannot
spread out freely from one source to the other and is confined in a vortex-like
configuration, which resembles the confinement of the magnetic field in
the Abrikosov vortex in the superconductor.

\section{Static Confining Potential}
It is straightforward to calculate the static confining potential using
the DGL lagrangian by putting a quark and an anti-quark separated by a
distance $r$.
\begin{equation}
\vec j_\mu (x)=\vec Qg_{\mu 0}\{\delta ^3({\bf x}-{\bf b})
-\delta ^3({\bf x}-{\bf a})\}.
\end{equation}
where ${\bf r}={\bf b}-{\bf a}$. This leads to the static potential as
\begin{eqnarray}
V(r)
&=&\vec Q^2 \int {d^3k \over (2\pi )^3}{1 \over 2}
(1-e^{i{\bf k}\cdot {\bf r}})(1-e^{-i{\bf k}\cdot {\bf r}})
\left [{1 \over {\bf k}^2+m_B^2}+{m_B^2 \over {{\bf k}^2 + m_B^2}}{n^2 \over
({{\bf n}\cdot {\bf k}})} \right ] \cr
&=&-{\vec Q^2 \over 4\pi }\cdot {e^{-m_Br} \over r}
+{\vec Q^2 m_B^2 \over 8\pi }\ln({m_B^2+m_\chi ^2 \over m_B^2})r \;,
\end{eqnarray}
besides the $r$-independent term for the case of type-II dual superconductor
($m_B < m_\chi$).
We get the finite expression because of the following the points,
; i.e. the direction of $\bf n$ to be parallel to
$\bf r$ ( ${\bf n} // {\bf r}$ ) due to the energy minimum condition and
the axial symmetry of the system, and the ultraviolet cutoff appears in the
perpendicular direction in $k_{_T}$ integration ($k_{_T} < m_\chi$)
according to the vanishing of the QCD-monopole field at the center of
the vortex-like configuration outward.
The Yukawa term provides a short range attraction
and the linear term a long range confining potential\cite{SST}.

The potential may be  adjusted to the phenomenological potential of the
Cornell group and the parameters are extracted by the fitting
procedure\cite{ALL}.
This procedure provides the masses of the glueballs, which are associated
with the properties of the QCD vacuum and quark confinement. It is very
important to find these glueballs with masses around 1GeV by experiment.

\section{Dynamical Chiral-Symmetry Breaking}
We would like now to discuss the behavior of quarks in the QCD-monopole
condensed
vacuum. This amounts to calculate the quark propagator using the Schwinger
-Dyson equation.
It seems, however, not straightforward, since quarks are confined.
When we want to describe a quark, there are every time other quarks nearby
and hence, in principle, we ought to treat the system of $q \bar q$ or $qqq$
or more complex ones with color singlet configurations totally.

This situation is similar to nuclear matter theory. In many-body nuclear
system,
each nucleon pair interacts strongly, in particular, the short range part of
the interaction even diverges.
Brueckner suggested first to solve a pair of nucleons, while the existence
of all the other nucleons is considered to block the quantum states
due to the Pauli effect.
Hence, a nucleon pair interacts through the $T$-matrix in free space,
\begin{equation}
T = V + V\;{1 \over e}\;T,
\end{equation}
while a nucleon pair in the many body system interacts through the $G$-matrix
with the Pauli exclusion operator $\hat Q$ as
\begin{equation}
G = V + V\;{{\hat Q} \over e}\;G.
\end{equation}
This Pauli exclusion operator makes the effective interaction $G$ moderate
in the medium. Then, we can make the Hatree-Fock procedure to calculate the
total energy and the single particle properties of nucleon in the medium.

Coming back to the properties of quarks, which are confined, we ought to
solve the Schwinger-Dyson equation,
\beqn
S^{-1}(p)= S_0^{-1}(p) +
\int_{0}^{\infty} S(p-q) D(q) dq \;\;,
\eeqn
which are written schematically. $S$ and $S_0$ denote the full propagator
and the simple propagator of a single quark, while $D$ the
gluon propagator. This expression does not take care of quark confinement,
however. Since quarks and gluons are confined inside of the hadrons,
we have to introduce a low momentum cutoff $q_c$, which is associated
with the size $R \sim q_c^{-1}$ of hadrons.
Hence, the SD equation ought to be modified to
\beqn
S^{-1}(p)= S_0^{-1}(p) +
\int_{q_c}^{\infty} S(p-q) D(q) dq \;\;,
\eeqn
\indent
Written rigorously, the Schwinger-Dyson equation in the DGL theory is
\beqn
S^{-1}(p)= i{p\kern -2mm /} +
\int {d^4q \over (2\pi)^4}
\vec Q^2
\gamma^\mu S(p-q) \gamma^\nu D_{\mu \nu}(q) \;,
\eeqn
assuming the following simple form for the full quark propagator,
\beqn
S^{-1}(p)= i {p \kern -2mm /}- M(p^2) \; .
\eeqn
We find the integral equation for $M(p^2)$ as
\beqn
M(p^2)= \int {d^4q \over (2\pi)^4} \vec Q^2
{M({(p-q)}^2) \over {(p-q)}^2+M^2({(p-q)}^2)} D_\mu^{\mu}(q^2) \; .
\label{quench}
\eeqn
where the trace of the gluon propagator in the Landau gauge is
\beqn
D^{\mu}_\mu(q^2)={2 \over q^2+m_B^2}+{1 \over q^2}
+{2 \over {(n\cdot q)^2+a^2}} \cdot
{m_B^2 \over q^2+m_B^2} \; .
\eeqn
Here, we have introduced a infrared cutoff $a$ ($\sim q_c$), for
the sake of numerical simplification, instead of the lower bound
of the integration region.

We take further the angular average on the direction of the Dirac string
because of light quarks moving in various directions inside of hadrons,
\beqn
\left\langle {1 \over {(n\cdot q)^2+a^2}} \right\rangle _{\rm average}
 \equiv {1 \over 2\pi ^2}\int {d\Omega_n}
          {1 \over (n\cdot q)^2+a^2}
={1 \over a} \cdot {2 \over {a +
\sqrt {a^2 + q^2} }} \; .
\label{kmave}
\eeqn
We find the SD equation,
\barr
M(p^2)
&=& \int_0^\infty {dk^2 \over 16\pi ^2}
{\vec Q^2 M(k^2) \over k^2+M^2(k^2)} \left({4k^2 \over
k^2+p^2+m_B^2+\sqrt {(k^2+p^2+m_B^2)^2-4k^2p^2}}
\right.
\nonumber \\
\jot 0.2cm
&&+{(1+\alpha _e)k^2 \over \max (k^2, p^2)}
     +{8 k^2 \over \pi a} \int_0^\pi d \theta
     {\sin^2 \theta \over a
     +\sqrt {k^2+p^2+a^2-2kp \cos \theta }} \nonumber \\
\jot 0.2cm
&& \times \left. \left[{m_B^2-a^2
    \over k^2+p^2+m_B^2-2kp\cos \theta }+{a^2
    \over k^2+p^2-2kp\cos \theta } \right] \right) \; .
\earr
where we have made the replacement, $k=p-q$.

We solve this SD equation numerically with the parameters ;
$e=5.5$, $m_B=0.5{\rm GeV}$ and $a=85{\rm MeV}$
with which the string tension is $k \simeq 1 {\rm GeV/fm}$.
We show in Fig.1 the quark mass $M(p^2)$ as a function of the
Euclidean momentum in unit of $\Lambda_{\rm QCD}$.
The glueball mass $m_B$ is proportional to the QCD-monopole condensate
and $M(p^2)$ increases with $m_B$.
This result indicates that the mechanism, which produces quark confinement,
is able to cause chiral symmetry breaking as well. Concerning the
condition for the parameters for the on-set of chiral symmetry breaking,
we find with some approximation, the following simple relation,
\beqn
e^2 m_B > 24\pi a \;.
\eeqn
This formula is very convenient to see that $m_B$ and hence
QCD-monopole condensation plays the essential role for chiral symmetry
breaking. Strictly speaking, the cutoff parameter $a$ depends on the glue
ball mass, $m_B$, because $a$ is closely related with the confinement effect.
With $m_B=0.5{\rm GeV}$, we find the quark mass, the pion decay
constant and also the quark condensate compare well with {\it experimental}
values.

\section{Recovery of Chiral Symmetry at Finite temperature}
The chiral symmetry is broken at zero temperature, which is nicely described
in the DGL theory. We then want to see if the theory provides the recovery
of the chiral symmetry as the temperature increases. We take the
imaginary-time formulation, and replace
\newpage
\barr
p_4 &\rightarrow& \omega_n = (2n+1)\pi T\nonumber \\
\jot 1.0cm
\int {{d^4 k} \over {(2\pi)}^4} &\rightarrow& T \sum_{m=-\infty}^{\infty}
\int {{{\rm d}{\bf k}} \over {(2\pi)}^3} \nonumber \\
\jot 1.0cm
M(p^2) &\rightarrow& M_{_T}(\omega_n ,{\bf p})
\earr
Hence, the SD equation at finite temperature is formulated as
\barr
M_{_T}(\omega_n,{\bf p})
&=& T\sum^{\infty}_{m=-\infty}
\int {{\rm d}{\bf k} \over (2\pi)^3} \vec Q^2
{M_{_T}(\omega_m,{\bf k}) \over {{\bf k}^2+\omega^2_m
+M^2_{_T}(\omega_m,{\bf k})}} \nonumber \\
&& \left[ { 2 \over { {\tilde k^2_{nm}} + m^2_B }  }
+ { 1 \over {\tilde k^2_{nm}} }
+ {4 \over a}{1 \over {a+
\sqrt{{\tilde k^2_{nm}}+a^2}}}
\left( { {m^2_B-a^2}
\over {{\tilde k^2_{nm}}+m^2_B} }
+ {a^2 \over {\tilde k^2_{nm}}}
\right) \right],
\earr
where ${\tilde k}^2_{nm}={(\omega_n - \omega_m)}^2+{({\bf p}-{\bf k})}^2$.
This form is too difficult to solve exactly. Hence, we take the
{\it covariant-like ansatz} for the self-energy at $T\neq0$ as
\beqn
M_{_T}(\omega_n ,{\bf p}) \simeq M_{_T}({\hat p}^2) \;\; {\rm with} \;\;
{\hat p}^2 = {\bf p}^2 + \omega_n^2
\eeqn
This assumption guarantees that the SD equation at $T\neq0$ reduces
to the one at $T=0$, when the angular average is taken in the three dimensional
momentum space.
We then find a manageable integral equation for
$M_{_T}({\hat p}^2)$.
We show in Fig.2 $M_{_T}({\hat p}^2)$ for various temperatures as a function of
${\hat p}^2$ in $\Lambda_{\rm QCD}$ unit. As expected, the mass function
$M_{_T}({\hat p}^2)$ decreases rapidly with temperature.
The quark condensate, $\langle {\bar q}q \rangle_T$,
shown in Fig.3, decreases gradually with temperature in the low temperature
region and vanishes suddenly near the critical temperature, in coincidence
with recent lattice QCD simulations\cite{karsh}.
The critical temperature is found around 110MeV. We should not take this
value seriously though. All the parameters in the DGL lagrangian may be
temperature
dependent. In addition, the infrared cutoff parameter $a$ should
be decreased with temperature, because the confinement effect gets weaker.
We should also take into account the change of $m_B$ for further
discussion of QCD phase transition at finite temperature\cite{ichie}.

\section{Conclusion}
We have studied chiral symmetry breaking in terms of the
dual Ginzburg-Landau (DGL) theory, which provides quark confinement.
We solve the SD equation by introducing infrared cutoff $a$, which is due to
the effect of the $q$-$\bar q$ pair creation and/or quark confinement.
We then have found that the DGL theory can provide chiral symmetry breaking
nicely with the parameters determined from the confining potential.
We have demonstrated also that the DGL theory provides the recovery of the
chiral symmetry at finite temperature.

We stress again that QCD-monopole condensation, formulated in the
DGL theory, causes both quark confinement and chiral symmetry
breaking. We mention also that QCD-monopole condensation is signaled by
the existence of glueballs with masses in the range of 1GeV ; $m_B(1^+)
\sim m_\chi(0^+) \sim 1{\rm GeV}$, and hope that these glueballs are found
experimentally, since they are the key particles for the confinement mechanism
due to QCD-monopole condensation\cite{matsubara}.

\section{Acknowledgment}
The authors are grateful to H.~Ichie, H.~Monden, T.~Suzuki and O.~Miyamura
for fruitful discussions on the QCD-monopoles.

\newpage

\end{document}